# Field-free perpendicular magnetization switching through domain wall motion in Pt/Co/Cr racetracks by spin orbit torques with the assistance of accompanying Joule heating effect

Baoshan Cui, Dong Li, Jijun Yun, Yalu Zuo, Xiaobin Guo, Kai Wu, Xu Zhang, Yupei Wang, Li Xi[*], and Desheng Xue

Key Laboratory for Magnetism and Magnetic Materials of Ministry of Education & School of Physical Science and Technology, Lanzhou University, P. R. China

**Abstract**

Heavy metal/ferromagnetic layers with perpendicular magnetic anisotropy (PMA) have potential applications for high-density information storage in racetrack memories and nonvolatile magnetic random access memories. Writing and erasing of information in these devices are carried out by domain wall (DW) motion and deterministic magnetization switching via electric current generated spin orbital torques (SOTs) with an assistance of in-plane bias field to break the symmetry. Improvements in energy efficiency could be obtained when the switching of perpendicular magnetization is controlled by an electric current generated SOTs without the in-plane bias fields. Here, we report on reversible electric-current-driven magnetization switching through DW motion in Pt/Co/Cr trilayers with PMA at room temperature due to the formation of homochiral Néel-type domain, in which an in-plane effective Dzyaloshinskii-Moriya interaction field exists. Fully deterministic magnetic magnetization switching in this trilayers is based on the enhancement of SOTs from a dedicated design of Pt/Co/Cr structures with two heavy metals Pt and Cr which show the opposite sign of spin Hall angles. We also demonstrated that the simultaneously accompanying Joule heating effect also plays a key role for field-free magnetization switching through the decrease of the propagation field.

[*] Corresponding author.   E-mail address: xili@lzu.edu.cn (Li Xi)

## I. Introduction

Recently, spin-orbit torques (SOTs) in sandwich structures with perpendicular magnetic anisotropy (PMA) where ultrathin ferromagnets (FM) is separated by a heavy metal (HM) and an oxide, have attracted abundant research interests for highly efficient magnetization switching[1-3] and fast domain wall motion.[4-10] In this kind of device, when an in-plane charge current ($J_e$) flows through HM with strong spin-orbit coupling (SOC) including Pt,[1, 11-13] β-Ta,[13-15] Hf,[16] and β-W,[17] etc., it can be converted into a pure spin current ($J_s$). Then, $J_s$ injects into FM and generates a torque to act on magnetic moments under the assistance of an in-plane magnetic field. As a result, if the torque is sufficiently strong, the magnetization could be switched. It is well established that the SOT switching efficiency is directly related to the magnitude of spin Hall angle ($\theta_{SH}$). So, considerable efforts have been devoted to obtain a large $\theta_{SH}$ of HMs by varying the thickness of HM,[18, 19] decorating the interface between HM and FM,[20, 21] changing the crystallinity of HM,[22] and even involving oxygen in HM.[23] Besides, some reports also achieve large effective $\theta_{SH}$ based on HM/FM/HM structures, in which two HM layers have opposite sign of $\theta_{SH}$.[24-26] However, a deterministic magnetization switching by SOT always requires an in-plane bias magnetic field along the current direction to break the symmetry,[1] which is indispensable to achieve the magnetization switching and has an obvious obstacle for the application in SOT-based devices. Aiming to realize field-free SOT devices, several designs through introducing an effective in-plane field have been experimentally demonstrated. The in-plane effective field could be induced via a wedge structure,[27, 28] or exchange coupling with another ferromagnetic layer with in-plane anisotropy,[29] exchange bias with an in-plane antiferromagnetic layer,[30, 31] and even in a hybrid ferromagnetic/ferroelectric structure.[32]

In our previous work,[33] we have investigated the deterministic magnetization switching in structural inversion asymmetric Pt/Co/Cr trilayer. In which the enhanced pure spin currents generated from both of Pt and Cr with opposite sign of $\theta_{SH}$ is used to work in concert to improve the SOT switching efficiency under an assistance of the in-plane magnetic field. The enhanced PMA and SOTs in Pt/Co/Cr structures were closely not only related to the Co/Cr, but to Pt/Co interfaces, where Dzyaloshinskii-Moriya interaction (DMI)[34-37] originates and shows influence on domain wall velocity as reported in recent works.[38-40] In our devices, there are two interfaces to contribute DMI, which could stabilize a homochiral Néel-type domain wall and result in an in-plane DMI effective field in DW. In this work, we report that the magnetization switching through current-driven domain wall motion also could be achieved in a micro-sized racetrack using the in-plane DMI field in a left-handed chirality Néel domain wall to replace an in-plane bias field. Moreover, we find that the accompanying Joule heating effect has a significant effect on current-driven domain wall motion and it can decrease the thermally activated barrier. Consequently, a small propagation field ($H_P$) is required to push DWM. While most works tend to ignore it[1, 14, 30] excepting few reports considering this effect.[41]

## II. Experimental

The Ta(3)/Pt(5)/Co(0.8)/Cr(1)/Al(1) (thickness number in nanometer) stacks with as-grown perpendicular magnetic anisotropy were deposited on Corning glass substrate by direct current magnetron sputtering with base pressure less than $5 \times 10^{-5}$ Pa as described elsewhere.[33] The deposited films were patterned into around 300 μm long and 8 μm wide racetracks with two circular nucleation pads and electrodes by photolithography and Ar ion milling as shown in Fig. 1(a). The typical resistance of the whole structure is around 6.3 kΩ. The formation of reversed magnetic domain and the creep of domain wall were observed by the variation of magnetic fields and/or the currents passing through the racetrack using a polar magneto-optical Kerr microscopy working in differential mode.[42] A voltage pulse generator (Tektronix: PSPL10300B, + 50/ -45 V with rising time around 300 ps) in a short pulse regime (15 − 100 ns) and wide period ranging from 10 μs to 1 s was used to generate pulse current to depin the DW and control the Joule heating effect. Current strength was calculated from the voltage measured in a serially connected real-time oscilloscope. A DC current source (Keithley: current source 220) was also used for comparison. All of the above experiments were carried out at room temperature.

**III. Results and discussion**

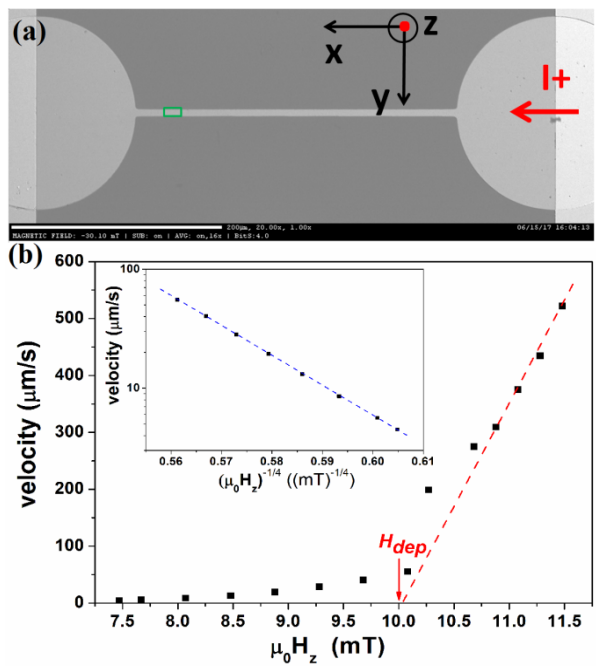

Fig. 1. Optical difference image of Pt/Co/Cr racetrack with two circular nucleation pads and electrodes saturated at -30.1 mT out-of-plane field (a) and domain wall motion velocity as a function of out-of-plane field $H_z$ (b). The green block in (a) shows the interested area, in which the hysteresis loops were recorded to obtain the propagation field at different magnetic fields and/or currents. The dash line in (b) is guided to the eye. The inset of (b) shows the scaling plot of log $v$ vs. $(\mu_0 H_Z)^{-1/4}$ with linear fitting.

The optical image of Pt/Co/Cr racetrack with two circular nucleation pads and electrodes is shown in Fig.1(a). In the experimental, we found that the thermal activated nucleation events will always happen in the left-side nucleation pads when the track was firstly saturated in one direction and then a small reverse magnetic field was applied. Thus, the domain wall (DW) propagates along the track from left side to right side driven by out-of-plane magnetic fields, i.e. the propagation field as shown in the supplement movies (SI-m1). The grey level of the interested green block area was recorded as a function of magnetic fields by the polar Kerr microscope working in the differential mode. Thus, the hysteresis loop can be obtained, from which the $H_P$ was obtained. The domain wall motion (DWM) velocity ($v$) was obtained by counting the time for DW travelling along the racetrack with a fixed length.

Figure 1(b) plots $v$ as a function of out-of-plane field $H_z$. The red arrow in the figure indicates the DW depinning field ($H_{dep}$), above which the DW exhibits a linear dependence with $v \propto H_z-H_{dep}$, which is the typical characterization of DW motion in the flow regime ($H_z > H_{dep}$).[43] Meanwhile, the creep regime ($H_z < H_{dep}$) shows thermally activated DWM with the creep criticality $\ln(v) \propto H_z^{-1/4}$,[43] as shown by the blue dash line of the best linear fit in the inset of Fig.2. In the following measurement, we will focus on the DW creep regime to investigate the current dependent DW velocity. In order to avoid too much Joule heating affecting the thermally activated barrier landscape, we use a pulse generator with pulse width lower than 100 ns and pulse period ranging from 10 μs to 1 s. We found that the variation of $v$ not only relies on the amplitude of current, but also is related to the polarity of the current.

When the pulse current is not more than 7.9 mA with the period of 10 μs, the nucleation events is always happened in their original position (the left nucleation pad) in this device. Moreover, the negative (positive) pulse currents favor the domain wall propagation along the track from the left (right) side to the right (left) side whenever the DW is up-down or down-up type since the nucleation always happens in the left side, and then DW always creeps from left to right along the track at a $H_P$. This kind of DWM characterization is shown in the supplementary movies (SI-m2). Thus, a negative current avails the DWM, while a positive current hinders the DWM along the track at a fixed $H_P$. This current direction determined DWM characterization can be ascribed to the enhanced SOTs from the Pt and Cr layers with opposite signs of spin Hall angles and the formation of chiral DW.[4, 15, 44] It should be mentioned that the current induced DWM by spin transfer torque[45] has a contrary effect from our experimental observation and has been ignored as reported in other SOT devices.[15, 44]

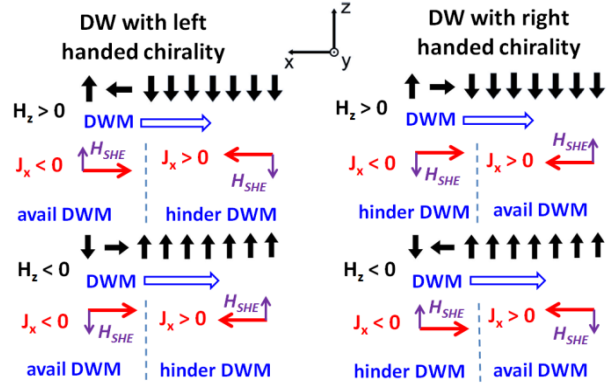

Fig. 2. The sketch of a left(right)-handed chirality Néel-type domain wall in Pt/Co/Cr sample with the illustration of negative current availed (hindered) or positive current hindered (availed) domain wall motion whenever the domain wall is up-down or down-up type. The anti-damping like spin Hall effective field ($H_{SHE}$) in all cases is also shown.

Figure 2 shows the sketch of domain wall motion under a pulse current and a propagation field $H_z$ with a left-handed and right-handed chirality DW. If there is only reversed $H_P$ applied, the DW will always creep from left side to right side as observed in our experimental. The pulse current generates a spin Hall effective field ($H_{SHE}$), which acts on the domain wall magnetization vector[4, 15, 44] and is given by,

$$\vec{H}_{SHE} = -\frac{\hbar \theta_{SHE} J_x}{2|e|M_s t_F}[\hat{m} \times (\hat{z} \times \hat{j})] \quad (1)$$

where, $\theta_{SHE}$, $M_s$, $t_F$, $J_x$, $\hat{m}$ and $\hat{j}$ represent the effective spin Hall angle, the saturation magnetization of the FM, the thickness of FM, the current density along x direction, the unit vector of magnetization of FM and the unit vector of current density, respectively. Current generated effective $H_{SHE}$ on a left (right) handed-chirality Néel DW is along (opposite) the external out-of-plane field $H_z$ when a negative current is applied, while it is opposite (along) to the direction of $H_z$ when a positive current is applied as shown in Fig.2. Thus, for a left-handed chirality DW a negative current induced $H_{SHE}$ will accelerate DWM, and a positive current induced $H_{SHE}$ will accelerate DWM for a right-handed chirality DW. In this work, the left-handed chirality of DW coincides with our experimental observations with the negative current availing the DWM and the positive current hindering DWM. Thus, a left-handed chirality DW forms in Pt/Co/Cr trilayers. It indicates a DMI in Pt/Co/Cr. The magnitude of DMI will be reported elsewhere.

The change of $H_P$ of the interested area with the variation of polarity and magnitude of the current occurs in Pt/Co/Cr with the specified left-handed chirality of DW. Figure 3(a) shows the normalized $H_P$ with variation of pulse current with the repeating frequency of 100 kHz, i.e. 10 μs periods. $H_P$ was obtained by measuring the hysteresis loop of the interested block area. One can see that the $H_P$ decreases with the decrease of the negative pulse current due to the negative pulse generated field has the same direction with the applied field during DW propagation, while at the positive current, $H_P$ firstly increases up to a peak around 5.8 mA and then decreases quickly. The increase of $H_P$ with the increase of positive pulse current can be ascribed to the increase of $H_{SHE}$, which is opposite to the applied field. The decrease of the $H_P$ at the high pulse current (e.g. 7.9 mA) may be ascribed to the Joule heating effect of current, which changes the thermal activated barrier landscape and makes the nucleation area changing to the left end of the track as shown in the supplementary videos (SI-m3). Moreover, the peak height increases with the decrease of pulse width at the same pulse period of 10 μs. It can be understood by that the shorter of pulse width is, the weaker Joule heating effect generates and the associated lower temperature rises, the slower DWM velocity becomes with the same strong spin Hall effective field, and the larger $H_P$ will be required to push the DW passing the interested block area, which leads to the increase of peak amplitude. When the pulse current is too large, the strong Joule heating effect will change the thermally activated barrier landscape and eventually changes the nucleation area and decreases the nucleation fields. It results in the quick decrease of $H_P$ at the large pulse current sides. These findings are also examined by keeping the pulse width in 100 ns and changing the periods to 100 μs (i.e. 10 kHz) as shown in Fig. 3(b). It can be seen that when decreasing the duty cycle, the decrease of $H_P$ on the large pulse current sides are all restrained due to the decreased Joule heating effect. Furthermore, we also examined our results by keeping the pulse number and pulse width are the same (i.e. the same SHE effect) in each magnetic field steps as shown in Fig. 3(c). In this condition, the function of SOTs is the same, but the Joule heating effect is larger for the shorter period one due to its higher duty cycle. It results in a large decrease of the $H_P$ due to the heating effect when the pulse current reaches 7.9 mA.

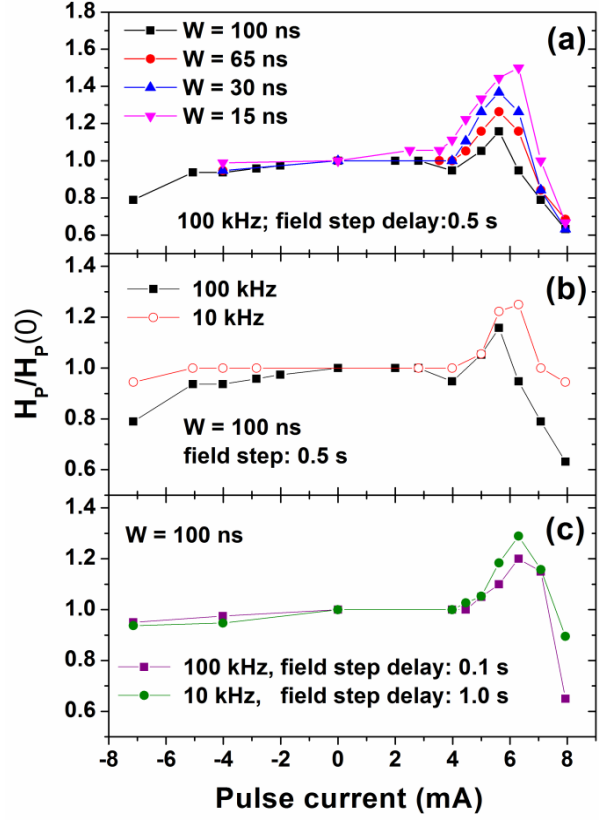

Fig. 3. Pulse current induced variation of normalized propagation fields with the same period and the different width of current pulses (a), and pulse current induced variation of normalized propagation fields with the same width and different periods of the pulse current and magnetic field steps (b, c). $H_P(0)$ represents the propagation field without any pulse currents.

The pulse current induced variation of the DWM velocity is also investigated with the fixed magnetic field and the variation of pulse current periods and magnitude as shown in Fig.6. In the creep region where $\mu_0 H_z <$ ~10 mT (see Fig. 1(b)), the DW velocity can be expressed as[43, 46, 47]

$$v = v_0 \exp\left[-\frac{U_C}{k_B T}\left(\frac{H_{dep}}{H}\right)^{1/4}\right] \quad (2)$$

where, $U_C$ is a characteristic energy related to the disorder-induced pinning potential, $k_B$ is the Boltzmann constant, $T$ is the temperature and $H_{dep}$ is the depinning field at which the Zeeman energy is equal to the DW pinning energy. At the same static out-of-plane magnetic field, $v$ is larger for the high frequency one due to the more heating generation, which increases $T$ in Eq. (2). Moreover, $v$ has quite large variation at the negative current side than that at the positive current side due to both of Joule heating effect and SOT effective field having the same contributions to $v$ at the negative current rather than their contrary contribution at the positive current as shown in Fig. 4(a). Eventually, the DW velocity can be changed by almost two orders of magnitude through negative pulse currents.

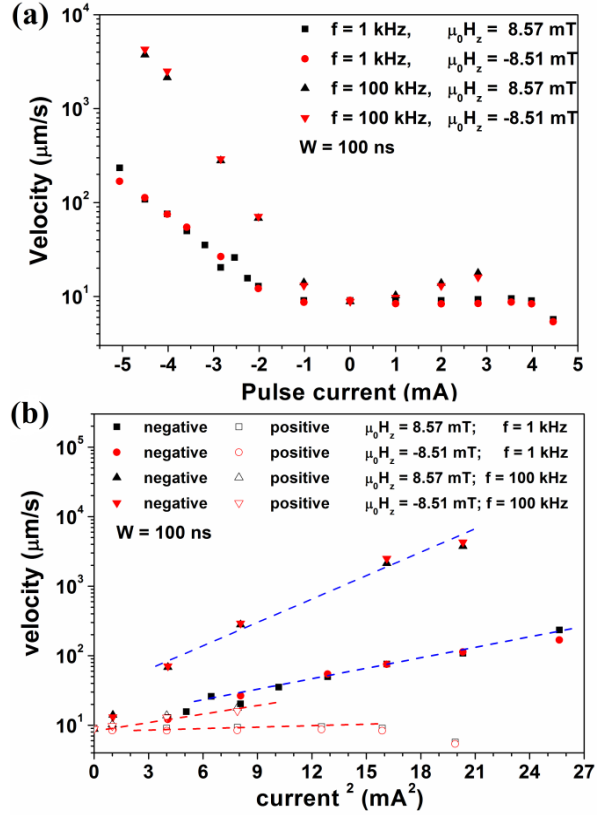

Fig. 4. The variation of DWM velocity with the different pulse currents under different magnetic fields and pulse periods(a), and the variation of DWM velocity with the square of pulse currents under different magnetic fields and pulse periods (b).

In our previous work, the effective antidamping-like SOT fields per unit current density is around 0.9 mT/($10^6$ A/cm$^2$) for Pt/Co/Cr sample.[33] Thus, the variation of pulse current corresponds to the variation of the effective $H_{SHE}$ acting on the DW. Thus we can quantitatively get the $v$ variation of DWM dependence on pulse current with a fixed reversal $H_P$ if the Joule heating effect does not exist. However, the above experimental already prove that the larger the pulse current becomes, the stronger the Joule heating effect will be. Thus, the current not only contributes to the variation of $T$, but also the spin Hall effective $H_{SHE}$ in Eq. (1). It is hard to distinguish those effects separately at this time. However, the Joule heating caused temperature rise is always proportional to $I^2$.[43] This temperature dependent effect was demonstrated in Fig. 4(b) which shows ln$v$ roughly proportional to the square of current. Moreover, the Joule heating effect becomes more evident if we change the pulse current to a DC current source (Keithley: 220). Figure 5 shows $H_P$ of the interested area in Fig. (1) at different DC currents. One can see the propagation field decreases with the increase of currents and is also roughly proportional to the square of the DC currents. Thus, the decrease of $H_P$ may be mainly ascribed to the Joule heating effect rather than the SOT effective fields. It should be mentioned that the nucleation position could be permanently changed from the left nucleation pad to the middle of the track due to thermal annealing effect once a quite high DC currents was applied (e.g. 3.2 mA in this case).

From the above observation, we can conclude that once a large current (no matter it is DC or pulse current) generated SHE effective field reaches the $H_P$, which was simultaneously reduced due to Joule heating effect of the current, the DW will pass the interested area to achieve the magnetization switching without the assistance of an in-plane field in our Pt/Co/Cr devices with left-handed chirality DW. Figure 6 shows snapshot of the DWM image at large current pulse around ±7.1 mA without any applied magnetic fields. In order to get strong $H_{SHE}$, we first use a reversed out-of-plane magnetic field to push the DW into the track, then the field was set to zero and a current pulse was applied. One can see DW was moved by the current pulse from left (right) side to right (left) side at a negative (positive) current pulse. The detailed DWM is shown in the supplementary movies (SI-m4), from which we can see that the reversible electric-current-driven magnetization switching through DW motion could be achieved via changing the polarity of the pulse current. Using the effective antidamping-like SOT fields per unit current density around 0.9 mT/($10^6$ A/cm$^2$) for Pt/Co/Cr sample with the assumption of uniformed flowing of current, the calculated effective antidamping-like SOT fields is around 8.2 mT, which is in the DW creep regime and can push the DWM. If only this $H_{SHE}$ existing, $v$ is around 10.5 μm/s according to Fig. 1(b), however, a quite large $v$ (>> 10 μm/s) is observed and ascribed to the Joule heating induced temperature rise, which on one hand could decrease the propagation field, and on other hand could increase $v$. It coincides with our other findings that the larger the pulse current with higher frequency, the larger the DWM velocity will be due to the current induced spin Hall effect and the accompanying Joule heating effect.

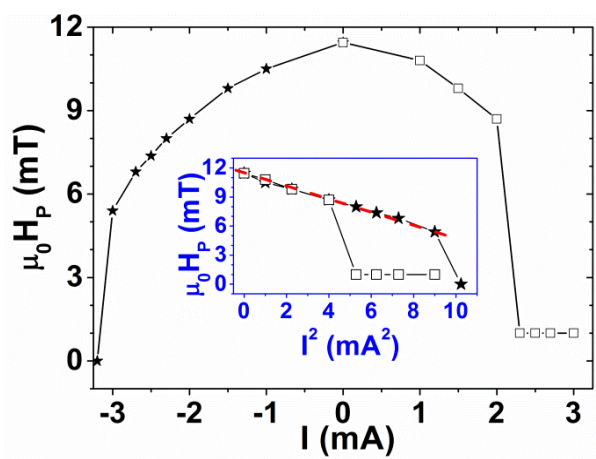

Fig. 5. DC current dependence of propagation field of the interested area. The inset shows the square of current dependence of propagation field. The dashed line is the best linear fitting curve.

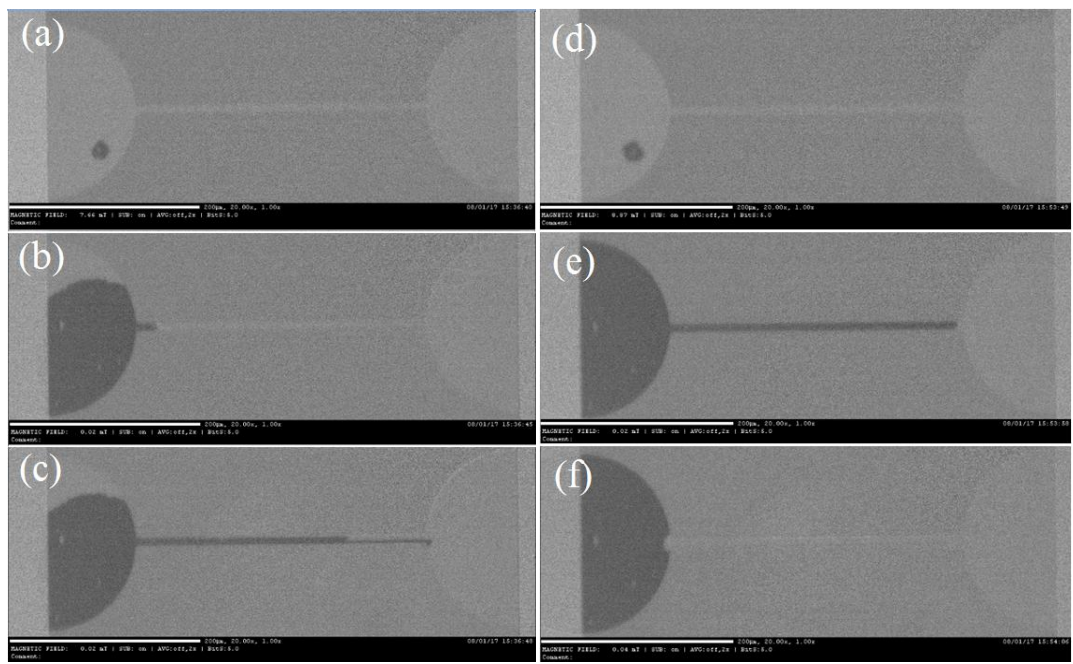

Fig. 6. Current pulse induced domain wall motion under negative current of -7.1 mA (a – c) and positive current of 7.1 mA (d – f). (a) and (d), the snapshots of the nucleation of a reversed domain. (b) and (e), the snapshots of DW, which was first pushed into the track by a reversed magnetic field, then the magnetic field was removed. (c) and (f), the snapshots of DW, which was pushed by turning on the corresponding current pulse.

In conclusion, the magnetization switching through domain wall motion in Pt/Co/Cr racetracks is achieved. The velocity of domain wall can be increased by more than two orders of magnitude through applying a negative current pulse by current generated extra spin Hall effective field and Joule heating effect in a fixed out-of-plane field comparing with that at a positive current pulse. Due to the formation of left-handed chirality domain wall, in which an in-plane effective DMI field exists, domain wall motion by current generated SHE effective field could be realized without applying any in-plane bias fields. The magnetization switching can be achieved by only applying a large current since the current generated spin Hall effective field can reach the propagation field, which was simultaneously reduced due to Joule heating effect of the current in our Pt/Co/Cr devices. The concept presented here will be useful in racetrack memories and logic devices and investigating the effects of SOT on the DW displacement.

## Acknowledgments

This work was supported by the Program for Changjiang Scholars and Innovative Research Team in University PCSIRT (No. IRT16R35), the National Natural Science Foundation of China (No. 51671098), the Natural Science Foundation of Gansu Province (No. 17JR5RA210) and the Fundamental Research Funds for the Central Universities (lzujbky-2015-122).

16